# Magneto-Moiré Excitons in Twisted Bilayer CrSBr

Qiuyang Li,[1,8,*] Anton Shubnic,[2,8] Nishkarsh Agarwal,[3] Adam Alfrey,[4] Wenhao Liu,[5] Zixin Zhai,[5] Igor Lobanov,[2] Valery Uzdin,[2] Senlei Li,[6] Yujie Yang,[1] Wyatt Alpers,[1] Kai Sun,[1] Liuyan Zhao,[1] Chunhui Rita Du,[6] Bing Lv,[5] Robert Hovden,[3] Ivan A. Shelykh,[2,7] Hui Deng[1,4,*]

[1]Department of Physics, University of Michigan, Ann Arbor, Michigan 48109, United States

[2]Department of Physics and Engineering, ITMO University, Saint-Petersburg 197101, Russia

[3]Department of Materials Science and Engineering, University of Michigan, Ann Arbor, Michigan 48103, United States

[4]Applied Physics Program, University of Michigan, Ann Arbor, Michigan 48109, United States

[5]Department of Physics, The University of Texas at Dallas, Richardson, Texas 75080, United States

[6]School of Physics, Georgia Institute of Technology, Atlanta, GA, 30332, United States

[7]Science Institute, University of Iceland, Dunhagi 3, IS-107, Reykjavik, Iceland

[8]These authors contributed equally to this work.

*Corresponding authors. Email: qyli@umich.edu; dengh@umich.edu.

**Moiré superlattices in van der Waals materials have revolutionized the study of electronic and excitonic systems by creating periodic electrostatic potentials. Extending this concept to magnetic materials promises new pathways in merging spintronics with photonics. While moiré magnetism has been revealed with near-field probes and nonlinear optical techniques, the coupling of these magnetic textures to optical excitations — magneto-moiré excitons — remains unexplored. Here, we report the observation of magneto-moiré excitons in twisted bilayer CrSBr, correlated with moiré spin textures that emerge below a critical twist angle of $\sim 2°$. The nanoscale moiré spin texture imprints distinct signatures onto the optical spectrum, shifting the exciton energy via a periodic magnetic exchange field. First-principles calculations corroborate that these signatures arise from one-dimensional spin textures governed by the balance of exchange interactions and domain wall energy. Our results demonstrate that moiré magnetism can be used to engineer nanoscale excitonic energy landscapes, providing a new platform for magneto-optical sensing, quantum transduction, and control of non-collinear magnetism and topology through light.**

Moiré superlattices formed in twisted van der Waals (vdW) heterostructures have opened a new frontier in emergent states of matter.[1–5] In non-magnetic semiconductors, such as transition metal dichalcogenides (TMDs), moiré superlattices lead to a periodic electrostatic potentials that trap excitons,[6–10] enabling programmable quantum emitters[11] and correlated quantum phases.[4,12–14] Integration of inherent spin orders[15] in moiré exciton systems may open a door to many exotic and emerging physical phenomena, yet magneto-moiré excitons have remained elusive.

A major challenge is the rarity of materials that simultaneously possess robust magnetic order, strong excitonic transitions, and potential to form moiré superlattices. Moiré magnetism has been observed in $CrI_3$[16,17] and $NiPS_3$[18,19] via near-field scanning probes and second harmonic generation symmetry analysis. However, excitonic resonances have not been identified in these moiré structures, likely due to fast dephasing or weak oscillator strength.

The layered antiferromagnet CrSBr[20] presents a promising system for exploring the interplay between moiré magnetism and moiré excitons, as it features excitons with exceptionally large oscillator strengths[21–24] and high binding energies.[25] Crucially, its optical properties are known to strongly couple to its interlayer magnetic order.[26–28] While a recent theoretical work has predicted novel magnetism in CrSBr moiré superlattices,[29] experimental realizations of twisted CrSBr bilayers have been confined to structures lacking moiré superlattices.[30–35]

Here, we report experimental realization of magneto-moiré excitons in twisted bilayer (tBL) CrSBr. By successfully fabricating samples with twist angles below a critical threshold of 2°, we observe distinct magnetic-field dependence of the exciton properties. We show that these features arise from the formation of quasi-one-dimensional (1D) moiré spin textures, which leads to a periodic modulation of the excitonic landscape. Our findings are supported by dark-field transmission electron microscopy (DF-TEM), which confirms the formation of rigid moiré lattices, and first-principles calculations, which reveal that the magnetic textures emerge in small-twist-angle moiré lattices due to competition between Heisenberg exchange interactions and domain wall formation. These results establish a unique platform where moiré excitons are modified by – and effectively probe – moiré spin textures, opening a door to novel magneto-optical quantum devices and optical manipulation of spin orders.

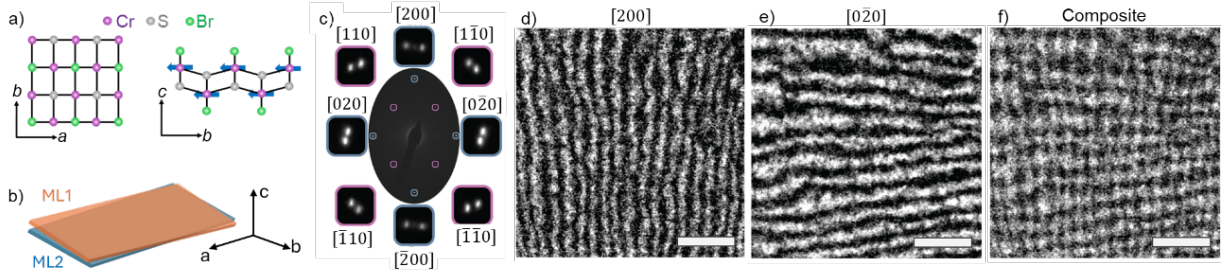

**Fig. 1. Moiré superlattice in CrSBr twisted bilayer (tBL). (a)** Scheme of the lattice structure of a CrSBr monolayer (ML) from the top (left panel) and side (right panel) views. Blue arrows: the spin direction in Cr atoms. **(b)** Scheme of CrSBr twisted bilayer (tBL). **(c)** Selected area electron diffraction (SAED) patterns of a 0.77°–CrSBr tBL at room temperature. **(d-e)** The dark-field transmission electron microscopy (DF-TEM) images from a selected area with electrons diffracted from (d) [200] and (e) [0$\bar{2}$0] SAED peaks. **(f)** The composite image of (d) and (e). Scale bars in (d-f): 50 nm.

**Formation of Moiré Superlattices in tBL CrSBr**

CrSBr monolayer (ML) is an anisotropic FM semiconductor with an orthorhombic lattice and magnetization along its crystalline b-axis (Fig. 1a), while CrSBr natural bilayer (nBL) is AFM with top- and bottom-layer spins anti-aligned along the c-axis due to the interlayer AFM exchange interaction.[26]

To form moiré superlattices, we stack two CrSBr MLs to create CrSBr tBLs with different twist angles $\theta$ (Fig. 1b) and measure the resulting bilayer structure via dark-field transmission electron microscopy (DF-TEM). The details are in Methods. Moiré superlattices are formed at small twist angles. As an example, Fig. 1c is the selected area electron diffraction (SAED) patterns of a CrSBr tBL, showing Bragg peaks with an out-of-plane two-fold rotation symmetry, corresponding to a local mean twist angle of $\theta = 0.77° \pm 0.09°$. Figs. 1d&e are the real-space images of the selected SAED region, constructed from two pairs of Bragg peaks with indices of [200] and [0$\bar{2}$0] showing moiré periods of 11.3±1.7 nm and 15.2±2.1 nm along a- and b-axis, corresponding to a twist angle of 0.87° ±0.12° and 0.88° ±0.12°, respectively. The composite image from the superposition of Figs. 1d&e is shown in Fig. 1f, which clearly presents the characteristic rectangular superlattice structures. The inhomogeneity of the moiré pattern could result from lattice strain and local structural disorders. The same measurements are performed on two additional tBL samples with $\theta \sim 1.4°$ and moiré superlattices are clearly observed in both (Supplementary Fig. 2 in Supplementary Materials). Consistent with the robust formation of a moiré superlattice at such small twist angles, the SAED pattern only shows two Bragg peaks without any observable satellite peaks corresponding to the lattice reconstructions. This is in contrast to other reported 2D vdW tBLs, such as graphene[36] and CrI$_3$,[37] and suggests that CrSBr has a more rigid lattice and is less prone to lattice reconstruction.

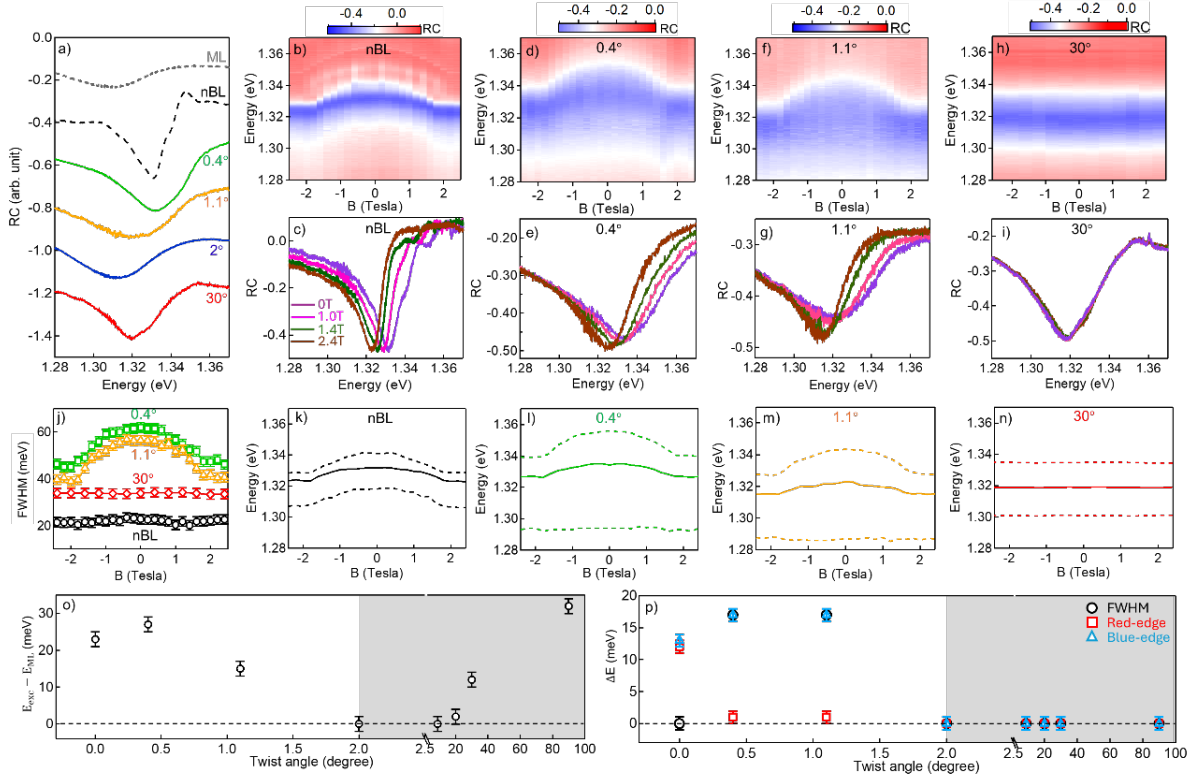

**Fig. 2. Moiré excitons in CrSBr tBLs and their dependence on twist angle and magnetic field.** **(a)** Comparison of reflection contrast (RC) spectra of CrSBr monolayer (ML), natural bilayer (nBL), and twisted bilayers (tBLs) with different twist angles. All curves are vertically shifted without normalization. Exciton absorption in tBLs shifts oppositely below and above $\theta = 2°$. **(b-i)** Magnetic field $B_\perp$ dependence of RC spectra of (b, c) nBL, (d, e) 0.4°–tBL, (f, g) 1.1°-tBL, and (h, i) 30°-tBL, in 2D plots for $B_\perp = -2.4T$ to 2.4T and 1D plots for $B_\perp = 0T$ (purple), 1.0T (pink), 1.4T (dark green), and 2.4T (dark brown). **(j)** Comparison of $B_\perp$ dependence of full width of half maximum (FWHM) of exciton peaks in nBL, 0.4°–tBL, 1.1°–tBL, and 30°–tBL. The FWHM of the nBL and 30°–tBL stay the same, but those of the 0.4°–tBL and 1.1°–tBL narrow significantly till saturation. **(k-n)** Comparison of the peak energy (solid lines), and red-side and blue-side energies of half-maximum of exciton peak (dashed lines) of (k) nBL, (l) 0.4°–tBL, (m) 1.1°–tBL, and (n) 30°–tBL. **(o)** Exciton energy difference between tBL (including nBL as 0°) and ML CrSBr as a function of twist angle. **(p)** Energy difference between $B_\perp = 0T$ and 2.4T as a function of twist angle for FWHM (black circles), red-edge (red squares) and blue-edge (blue triangles) at the half maximum of exciton peak for tBLs (including nBL as 0°). With increasing $B_\perp$, spectra of the nBL red shift but keep the same lineshape, of the 0.4°–tBL and 1.1°-tBL both red shift and narrow significantly, and of the 30°-tBL don't change.

**Spectral Signatures of Magneto-Moiré Excitons**

We then characterize the optical properties of the CrSBr tBLs of different twist angles, in comparison with ML and nBL, by magneto reflection contrast (RC) spectroscopy at a temperature of 5 K, under an out-of-plane magnetic field $B_\perp$ from zero to $\pm 2.4$ T, for a saturation field of about 1.8 T, as presented in Fig. 2. Evident in Fig. 2, while all the samples show the characteristic negative peak of the magnetically confined surface excitonic state,[24] tBLs with $\theta < 2°$ show distinct properties compared to that of ML, nBL, and tBL with $\theta \geq 2°$, in regards to the exciton energy and line shape (Fig. 2a), and magnetic field dependence of both the energy (Figs. 2b-i) and linewidth (Figs. 2j-n) of the exciton.

Fig. 2a compares the RC spectra of the different samples at $B_\perp = 0$. The ML has a broad exciton absorption with a peak position of 1.306±0.002 eV; in the nBL, the exciton is blue-shifted by 26 meV due to interlayer coupling and orbital hybridization that leads to an AFM order.[26–28] For tBLs with $\theta \geq 2°$, the exciton blue-shifts from close to the ML exciton energy and the exciton oscillator strength is about twice of that of the ML; this has been attributed to the interlayer coupling becoming negligible while the decrease in overall dielectric anisotropy leads to reduced exciton binding energy.[35] These results are fully consistent with previous reports in the literature. Interestingly, however, for tBLs with a very small twist angle, as $\theta$ increases from 0.4° to 2°, the exciton peak redshifts instead from that of the nBL to that of ML, before it starts to blueshift for $\theta \geq 2°$. This result suggests that there is a finite interlayer coupling in tBL with $\theta < 2°$, similar to that in nBL but decreasing with $\theta$.

The finite interlayer coupling in small–$\theta$ tBL is further confirmed in the $B_\perp$ dependence of the exciton energy, as shown in Figs. 2b-i. In ML and large–$\theta$ tBL ($\theta \geq 2°$), the exciton resonance does not change with $B_\perp$, as shown by the example of 30°-tBL in Figs. 2h-j. In contrast, in nBL (Figs. 2b-c) with strong interlayer electronic and magnetic coupling, the exciton energy redshifts continuously with increasing $B_\perp$, by up to 9 meV till saturating at 1.8 T, as the spin order changes from in-plane AFM to fully out-of-plane FM.[26] Similar to the nBL, in the 0.4°–tBL and 1.1°–tBL (Figs. 2d-g), exciton energy redshift with increasing $B_\perp$, by up to 12 meV and saturating at 1.8 T. This result confirms that, unlike the larger-$\theta$ tBLs but similar to the nBL, the spin orders in the two monolayers in 0.4°– and 1.1°–tBLs are strongly coupled with each other.

Distinct from nBL, however, the 0.4°– and 1.1°–tBLs feature significant change in the exciton linewidth and line shape under $B_\perp$ (Figs. 2j-n). As shown in Figs. 2j-n and summarized in Fig. 2p, the full width at the half maximum (FWHM), as well as the line shapes, of the excitons in both the nBL and 30°-tBL change negligibly with $B_\perp$. In stark contrast, the FWHM of the 0.4°– and 1.1°–tBLs narrow by 16±4 meV as $B_\perp$ is increased from 0T to ≥1.8T (Fig. 2j and p), and, importantly, the decrease mainly results from higher-energy components red-shifting and merging with the lower energy ones while the lower energy edges stay unshifted (Figs. 2l-m and p). This unique $B_\perp$-dependence of the linewidth and line shape cannot be explained by a uniform magnetic order. Instead, it shows that spin alignment varies in these tBLs and exciton properties are strongly correlated with the resulting magnetic domains. Excitons in FM like domains have lower energies and do not change under $B_\perp$; excitons from AFM like domains have higher energies, and they redshift and merge with the lower energy ones as $B_\perp$ changes the local spin order from AFM to FM.

In short, as summarized in Figs. 2o-p, tBLs show distinct behaviors below or above the critical angle of $\theta = 2°$. As shown in Fig. 2o, below 2°, tBL exciton energy redshifts from nBL, expected from finite but reduced interlayer electronic coupling; above 2°, tBL exciton energy blueshifts instead, due to twist-dependent dielectric anisotropy rather than interlayer tunneling.[35] Consistently, as shown in Fig. 2p, there is a strong $B_\perp$ dependence of the FWHM and higher-energy contributions to exciton absorption below 2°, due to the interlayer electronic and magnetic coupling; while no $B_\perp$ dependence is observed above 2°. These observations suggest interlayer electronic and magnetic coupling that leads to the formation of moiré magnetic domains in tBL with $\theta < 2°$, and furthermore, moiré exciton resonances correlated with these local domains.

Formation of spin textures and magnetic domains within the moiré unit cells has been observed in $CrI_3$ twisted bilayers[38,39] as well as double-bilayers[37,40] and double-trilayers,[38,41] but excitonic transitions have not been observable in these systems. Recently, magnetic domains have been theoretically predicted in CrSBr tBLs.[29] Our observations suggest that, unique to CrSBr, strong local exciton resonances survive in the moiré superlattice and, most interestingly, their properties are strongly correlated with the spin ordering, allowing easily accessible optical probes of the moiré magnetic order.

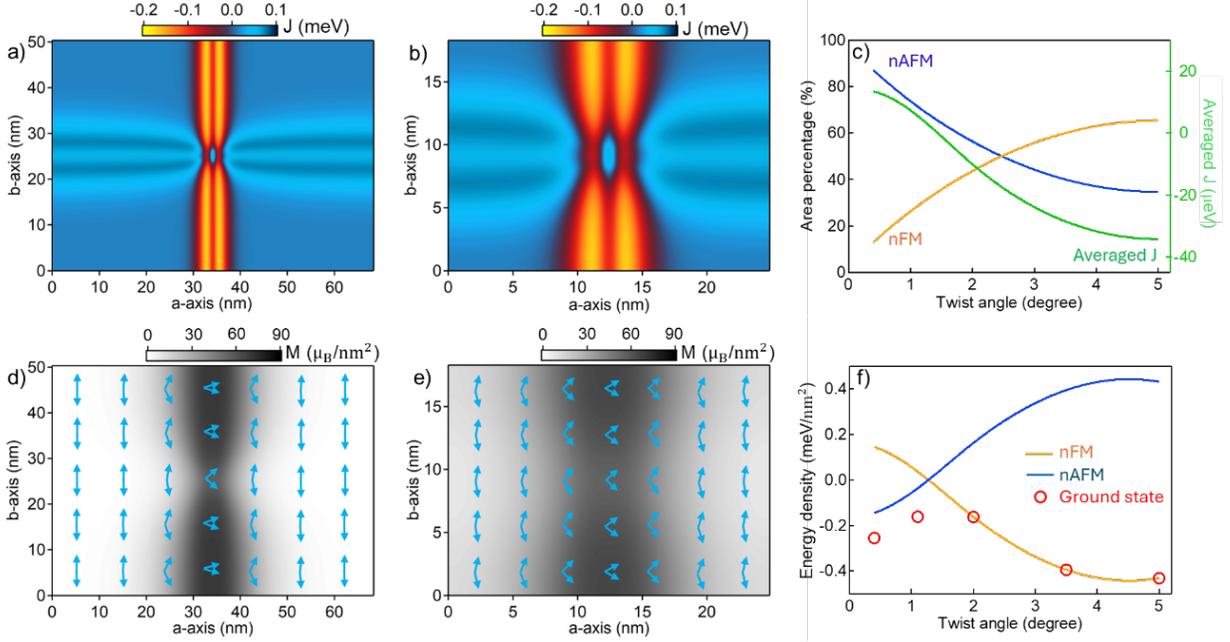

**Fig. 3. Calculated magnetic energy and moiré spin textures in CrSBr tBLs. (a-b)** Maps of interlayer exchange coupling strength (J, color intensity) within a moiré unit cell for (a) 0.4°–tBL and (b) 1.1°–tBL, showing $J < 0$ (nFM favored) and $J > 0$ (nAFM favored) regions. **(c)** Twist angle dependence of the fractional area of $J < 0$ (orange line) and $J > 0$ (blue line) regions (left y-axis), and averaged $J$ over a moiré unit cell (green line, right y-axis). The system becomes more nFM favored with increasing twist angle. **(d-e)** Maps of magnetization (M, color intensity) and spin orientation distribution (blue arrows) for (d) 0.4°–tBL and (e) 1.1°–tBL, showing smoothly evolving spin textures and net magnetizations along a-axis. **(f)** Twist angle dependence of the energy densities of nFM (orange line), nAFM (blue line), and the true ground state (red circles). The ground state is uniform nFM for $\theta > 2°$; but it differs from both nFM and nAFM for $\theta = 0.4°$ and 1.1°, featuring spin textures shown in (d) and (e).

**Theoretical Analysis of Moiré Spin Texture Formation**

We further confirm the formation of quasi 1D magnetic domains and correlation with exciton properties by *ab initio* calculations of CrSBr tBLs with different twist angles within the density functional theory (DFT) framework (see Methods for details). Figs. 3a&b show the calculated maps of interlayer exchange coupling strength, $J$, in the moiré unit cell of the 0.4°–tBL and 1.1°–tBL, assuming the spins in each monolayer remain aligned with the respective b-axis. There are regions with positive and negative $J$, which favor interlayer alignment and anti-alignment of the spins, respectively. Due to the finite twist angle, they correspond to non-collinear magnetic orders that are more FM like (nFM) and more AFM like (nAFM), respectively. For twist angles between 0.4° and 5°, the average $J$ within a moiré unit cell decreases from 13.4 to -34.3 μeV (Fig. 3c, right axis); correspondingly, the fractional area of nFM ($J > 0$) favored region increases from 12.9% to 65.3%, while the nAFM region decreases from 87.1% to 34.7% (left y-axis). These results suggest that nFM order becomes more favored with increasing twist angle, consistent with previous calculations.[29] The global magnetic ground state depends on both the spatial distribution of $J$ and energy cost of domain walls. Taking into account domain walls in the calculation, we obtain the global magnetic ground state of tBLs of different twist angles. Figs. 3d-e show the maps of the calculated magnetizations along the a-axis (M, color intensity) and corresponding spin orientations (blue arrows) for 0.4°– and 1.1°–tBLs, respectively. More detailed maps are shown in Supplementary Fig. 6 in Supplementary Materials (SM). These results show that, in 0.4°– and 1.1°–tBLs, the spin alignment evolves continuously over space, rotating symmetrically from their respective b-axes toward or away from each other, forming varying net magnetizations along the a-axis and quasi 1D feature.

Fig. 3f shows the energy density of the global magnetic ground state (red circles), compared to those of nFM (orange line) and nAFM (blue line) states. For $\theta$ <2°, both nFM and nAFM have higher energy densities than the calculated ground state with spatially varying spin alignment (Figs. 3d&e). However, for $\theta$ >2°, the ground state becomes nFM ordered. This is because the energy cost of the domain wall scales as $\sqrt{A}$ for moiré unit cell area $A$, while that of nAFM *vs.* nFM scales as $A$. As moiré unit cell becomes smaller, the energy cost of domain walls becomes higher than the energy saving of small nAFM domains, and a single domain of the lower energetic nFM order becomes preferred. The critical twist angle of about 2°, corresponding to a critical unit cell area of 250 nm$^2$, matches well the experimental observations.[35]

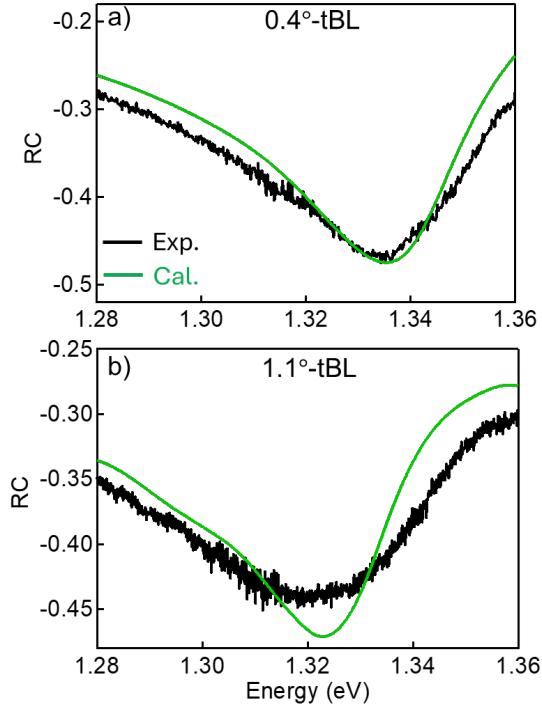

**Fig. 4. Correlations between moiré excitons and moiré magnetisms.** The experimentally measured (Exp., black lines) and calculated (Cal., green lines) RC spectra at 0T of (a) 0.4°–tBL and (b) 1.1°–tBL.

**Correlation between Moiré Exciton Spectra and Moiré Spin Texture**

To understand the correlation between the moiré exciton spectral properties and moiré magnetisms, we compare the measured RC spectra at 0T of 0.4°–tBL and 1.1°–tBL with calculated ones based on moiré spin textures (green lines in Fig. 4). In ML, nBL and large-$\theta$ tBLs, where there is a uniform magnetic order without multiple domains, the exciton line shape is independent of $B_\perp$; in nBL, the exciton energy redshifts when the uniform magnetic order is changed from AFM to FM, as observed in our sample (Fig. 2) and reported in the literature.[26] Based on this, we consider that excitons at different local domains of a small-$\theta$ tBL all have the same line shape but have different center energies determined by the local interlayer spin alignment. The line shape can be obtained by RC spectra at 2.4 T, above the saturation field where a single FM domain is formed, with all spins anti-aligned with $B_\perp$. We then shift the spectra at 2.4T based on the interlayer spin alignment map (Figs. 3d&e and Supplementary Fig. 6 in SM) and average over the unit cell to construct the RC spectra at 0T (see Methods for details). The maximum spectral shift of tBL from AFM to FM ordered states ($\Delta E$) is the only fitting parameter and is determined to be 12.5 meV for both 0.4°– and 1.1°–tBLs. The resulting spectra for 0.4°– and 1.1°–tBLs are shown in Fig. 4, which agree well with the measured spectra, providing further support of the spin texture and its correlation with exciton spectral properties. The agreement is better for the 0.4°–tBL than the 1.1°–tBL. The deviation could result from the spatial size of the exciton. Our current model neglects the effect of the size of exciton and shifts the exciton energy on the length scale of spin alignment variation; the latter is reduced with increasing $\theta$, resulting in potentially a greater error for larger $\theta$. Further studies of the band structure and exciton wavefunction of CrSBr tBLs may provide a more accurate prediction of the exciton spectrum.

**Discussion and Outlook**

In summary, we have demonstrated the realization of magneto-moiré excitons in twisted bilayer CrSBr, revealing a strong correlation between excitonic resonances and quasi-one-dimensional moiré spin textures. Our results identify a critical twist angle of about 2°; in moiré superlattices with smaller twist angles, the competition between interlayer exchange interactions and domain wall formation energy drives the emergence of complex magnetic textures. Unlike many vdW heterostructures that undergo lattice reconstruction, the structural rigidity of CrSBr preserves the moiré lattice even at these small twist angles, enabling the formation of stable, continuous spin domains that evolve between antiferromagnetic and ferromagnetic alignments.

These findings establish small-angle twisted CrSBr as a unique platform where moiré magnetism creates an excitonic energy landscape, which, in turn, allows for the direct optical readout of nanoscale spin textures, bypassing the need for invasive local probes. This platform opens a versatile pathway for exploring emergent quantum phases, such as moiré domain wall magnons[42] and Hubbard excitons in AFM domains.[43] The ability to imprint magnetic potentials onto optical excitations lays the foundation for a new generation of magneto-optical technologies, ranging from quantum transduction and sensing to the programmable control of non-collinear magnetism via light.

**Data availability**

All data that support the results of this article are available from the corresponding authors upon request.

## Methods

### Growth of CrSBr single crystals

We grew CrSBr single crystals with the solid−vapor method in a box furnace following the reported procedures.[44] Typically, we loaded Cr powders (Alfa Aesar, 99.97%), S powders (Alfa Aesar,99.5%), and solid $Br_2$ (solidified from liquid bromine, Alfa Aesar 99.8%, with assistance of liquid nitrogen) with the mole ratio of 1:1.1:1.2 in the clean quartz ampoules inside an argon glovebox. We flame-sealed the ampoules under vacuum with the liquid nitrogen trap. The sealed ampoules were then heated up to 930 °C slowly and kept there for 20 h, cooled down to 750 °C at a rate of 1°C/h, and finally quenched down to room temperature. CrSBr single crystals formed at the bottom of the ampoules, which were well separated from byproducts of $CrBr_3$ that formed at the top of the ampoules.

### Fabrication of CrSBr tBLs

The whole sample fabrication process was done inside a nitrogen-glovebox. We first exfoliated CrSBr flakes and transferred them onto 90 nm $SiO_2$/Si substrates (0.5 mm thick) using scotch tape and identified the CrSBr monolayer (ML) by their color with a fixed microscope setting. The CrSBr's color-thickness relationship was confirmed by atomic force microscopy as reported previously.[35] We then fabricated CrSBr twisted bilayers (tBLs) by transferring one CrSBr ML onto another with a polymer film (1.2g poly(bisphenol A carbonate) dispersed in 20g chloroform). The twist angles were controlled by aligning the long crystal axis (a-axis) of the flakes. For TEM measurements, we transferred tBL onto 10 nm $Si_3N_4$ membranes. For optical measurements, we transferred tBLs onto 90 nm $SiO_2$/Si substrates and the twist angles were confirmed by polarization-dependent reflection contrast (RC) measurements on each ML as shown in Supplementary Fig. 1 in SM. The normalized exciton peak intensity at 1.306 eV for each ML in tBL samples were plotted as function of polarization angle ($\alpha$). These polarization dependent curves for each ML were fitted by $\cos^2(\alpha-\varphi_i)$, where $\varphi_i$ (i=1, 2) is the relative angle of ML1 and ML2, respectively. Twist angle was then determined as $\Delta\varphi=|\varphi_1 - \varphi_2|$.

### Optical measurements

We conducted reflection contrast (RC) spectroscopy by real-space imaging of the samples. For measurements without magnetic fields, we kept the samples at 5 K with a Montana Fusion system. We used an objective lens with numerical aperture (NA) for both focusing and collection. For magnetic field dependent measurements, we kept the sample at 4.5 K inside an AttoDry 1000 system and we used an objective lens with NA of 0.90 for both focusing and collection. We used a supercontinuum white light laser (NKT Photonics, superK) with a beam size of ~2 μm in diameter as the white-light source for all RC measurements. The signals were detected by a Princeton Instruments spectrometer with a cooled charge-coupled camera. Polarization-dependent RC spectra were collected by rotating the polarization orientation of a linearly polarized white light with a half waveplate and recording the RC intensity as a function of the rotation angle.

**SAED and DF-TEM measurements**

We used the diffraction-based transmission electron microscopy (TEM) measurements to characterize moiré superlattice structures of CrSBr tBLs. Thermo Fisher Talos, operated at 200kV and equipped with Gatan OneView camera, was employed for SAED and DF-TEM imaging. The local mean twist angle was measured by fitting 2D gaussians to the Bragg peaks. The composite DF-TEM real space image was obtained in post-processing by averaging DF-TEM images from two third order [200] Bragg peaks 90° apart.

**Modeling of magnetic structures in CrSBr tBLs**

The magnetic properties of CrSBr tBLs with a small twist angle (<5°) are strongly influenced by the effect of lattice reconstruction. We first calculated this effect using a multiscale modeling method proposed in ref[45]. Ab initio calculations were performed within the density functional theory (DFT) framework using the GPAW package.[46,47] In all calculations, we used experimental values of the bulk CrSBr lattice constants a = 3.51 Å and b = 4.77 Å.[48] We used the optB88-vdW functional with a plane-wave energy cutoff of 900 eV as implemented in libvdwxc library.[49] The Brillouin zone was sampled by a 8×8×1 k-point mesh and a 15 Å thick vacuum layer was added in c-axis direction to avoid spurious interactions. The positions of the atoms in the CrSBr monolayer unit cell were optimized using the Broyden—Fletcher—Goldfarb—Shanno (BFGS) algorithm.

We approximated the CrSBr bilayer adhesion energy W(**R**, d) on the mutual shift **R** of the top and bottom monolayers relative to each other (stacking) and the interlayer distance d by the equation:

$$W(\mathbf{R}, d) = W_0(d) + \sum_i W_i(d) \cos(G_{ix} R_x) \cos(G_{iy} R_y) \tag{1}$$

where $G_{1-6}$ are the 6 smallest vectors of the reciprocal lattice: $G_1 = G_a$, $G_2 = G_b$, $G_3 = 2G_a$, $G_4 = 2G_b$, $G_5 = G_a + G_b$, $G_6 = G_a + 2G_b$, where $G_a = \left(\frac{2\pi}{a}, 0\right)$ and $G_b = \left(0, \frac{2\pi}{b}\right)$. We performed the DFT calculations of the $W(\mathbf{R}, d)$ on a 4×4 **R**-space grid for a series of interlayer distances. Then the $W(\mathbf{R}, d)$ values were fitted using Eq. (1) and the $W_i(d)$ were obtained by cubic spline interpolation. Our DFT calculations show that $W(\mathbf{R}, d)$ is symmetric with respect to reflection relative to the CrSBr's crystallographic axes a and b. Thus, only the terms with cosines appear in the Eq. (1). We expanded all the terms in Eq. (1) near the $W_0(d)$ minimum and obtained the $d(\mathbf{R})$ by minimizing $W(\mathbf{R}, d)$ with respect to d:[50,51]

$$d(\mathbf{R}) = d_0 - \frac{1}{2\varepsilon} \sum_i W_i'(d_0) \cos(G_{ix} R_x) \cos(G_{iy} R_y) \quad (2)$$

where $W_0(d) \approx \varepsilon(d - d_0)^2$. The local stacking of bilayers twisted at a small angle θ can be calculated as $\mathbf{R} = \mathbf{u}(\mathbf{r}) + \theta \hat{z} \times \mathbf{r}$, $\mathbf{u}(\mathbf{r})$ is the local displacement of the monolayers relative to each other. $\mathbf{u}(\mathbf{r}) = \mathbf{u}^t(\mathbf{r}) - \mathbf{u}^b(\mathbf{r})$, where $\mathbf{u}^t(\mathbf{r})$ and $\mathbf{u}^b(\mathbf{r})$ are the local displacements of the top and bottom monolayers relative to their non-relaxed positions at the point **r**, respectively. $\mathbf{u}(\mathbf{r})$ can be found by minimizing the sum of the adhesion energy and the elastic energy resulting from deformation:

$$W = \int d^2\mathbf{r} \left[ \sum_{k=t,b} (\mathbf{u}^k \hat{C} \mathbf{u}^k) + \varepsilon(d(\mathbf{r}) - d_0)^2 + \sum_i W_i(d_0) \cos(G_{ix} R_x) \cos(G_{iy} R_y) \right] \quad (3)$$

Here $\hat{C}$ is the elasticity tensor. We calculated the CrSBr elasticity tensor using the finite deformation method with a PBE exchange-correlation functional[52] as implemented in the Elastic package[53] and obtained the following nonzero components of the elasticity tensor ($\hat{C}$): $C_{11} = 115$ N/m, $C_{11} = 111$ N/m, $C_{12} = 16$ N/m, $C_{66} = 22$ N/m. The footnotes (i, j = 1, 2, 6) of the component $C_{ij}$ indicate the displacement tensor matrix elements, where 1 represents $\mathbf{u}_{xx}$, 2 represents $\mathbf{u}_{yy}$, and 6 represents $\mathbf{u}_{xy}$.

We reduced the problem of minimizing W to the problem of solving the Euler-Lagrange equations with periodic boundary conditions. The finite difference method on a rectangular grid was used. To solve the system of nonlinear algebraic equations, we used the interior point method, as implemented in the GEKKO Optimization Suite.[54] The spatial distribution of $u_y(\mathbf{r})$ are shown in Supplementary Fig. 3 in SM. The upper panels demonstrate that the lattice relaxation of the 5.0°– tBL leads to a small modification of the moiré unit cell geometry. However, the adhesion energy term increases with decreasing twist angle, and in the 0.4°–tBL, well-defined domains with a domain wall between them are formed. Similar results for $u_x(\mathbf{r})$ are shown in Supplementary Fig. 4 in SM.

To study the magnetic configuration of the CrSBr tBL, we calculated a map of the interlayer magnetic exchange constant spatial distribution in the moiré supercell. We calculated the interlayer magnetic exchange constant J on a 4×4 real-space displacements grid and approximated the $J(\mathbf{R})$ by the equation:[55]

$$J(\mathbf{R}) = J_0 + \sum_i J_i \cos(G_{ix} R_x) \cos(G_{iy} R_y) \qquad (4)$$

where $\mathbf{G_1} = \mathbf{G_a}$, $\mathbf{G_2} = \mathbf{G_b}$, $\mathbf{G_3} = 2\mathbf{G_a}$, $\mathbf{G_4} = 2\mathbf{G_b}$, $\mathbf{G_5} = \mathbf{G_a} + \mathbf{G_b}$. Ab initio calculations were performed using the Quantum ESPRESSO package.[56] The LDA+U+J approach[57] with U=0.5 eV, J=0.03 eV for 3d Cr electrons was used. We selected the values of U and J in such a way as to reproduce the value of the CrSBr nBL interlayer exchange magnetic constant.[34] The van der Waals interaction was taken into account by the DFT-D2 method. A plane-wave energy cutoff of 120 Ry and a dense 16×16×1 k-point mesh were used.

Supplementary Fig. 5 in SM shows the $J(\mathbf{r})$ maps calculated based on the $u_x(\mathbf{r})$ and $u_y(\mathbf{r})$ maps. The $J(\mathbf{r})$ map of the 5.0°–tBL without lattice relaxation demonstrates the dependence of J on the displacement $\mathbf{R}$. It is clearly seen that enhancing lattice relaxation with decreasing twist angle leads to a decrease in the area of regions with nFM type of ordering. Simultaneously, the $J(\mathbf{r})$ modulus in the nFM regions is significantly larger than in the nAFM ones, which results in a homogeneous nFM ordering of the 5.0°– and 2.0°–tBL. In the 1.1°–tBL, the area of the nAFM region becomes large enough to change the most energetically favorable homogeneous magnetic ordering from nFM to nAFM. However, due to the high $J(\mathbf{r})$ modulus values in nFM regions, the formation of a domain structure becomes possible.

To perform a simulation of the CrSBr tBL magnetic structure, we used the Magnes package,[58] which has been successfully used to model the domain structure of moiré magnets.[59] We calculated the magnetic ground state within the framework of the micromagnetic model,[60] considering intralayer and interlayer exchange interactions as well as single-ion anisotropy. The constants of intralayer exchange and single-ion anisotropy were taken from ref.[61] The simulations confirmed the formation of magnetic domains. tBLs with twist angles of 2° or more are characterized by a homogeneous nFM ordering. The magnetization maps for the 1.1° and 0.4° twist angles are given in Fig. 3 and Supplementary Fig. 6 in SM.

**Modelling of RC spectra of tBLs at 0T**

We reproduce the RC spectra of tBLs at 0T based on their RC spectra at 2.4T and the calculated magnetic structures. We first divided the points in the RC spectra at 2.4T into groups of 50 and averaged the values in each group to get a smooth curve. Then we interpolated the resulting curves with cubic splines and got the processed RC spectra at 2.4T, $S_{2.4T}(E)$, as shown in Supplementary Fig. 7 in SM, which represents the spectra of the single FM ordered exciton state. Exciton energy of CrSBr bilayer is sensitive to the twist angle between top and bottom layer spins (β) and shifts as: $\Delta E \cdot \cos^2(\beta/2)$, where $\Delta E$ is the maximum energy shift when β changes from 180° to 0° (from AFM to FM order).[27] We first reproduced the RC spectra at 0T, $S_{0T}(E)$, by considering excitons in each CrSBr atomic unitcell separately within a moiré unit cell:

$$S_{0T}(E) = \frac{\sum_{x,y} S_{2.4T}\left(E - \Delta E \cdot \cos^2\left(\frac{\beta_{x,y}}{2}\right)\right)}{N_{x,y}} \quad (5)$$

where the summation in the numerator is over the moiré unit cell, $\beta_{x,y}$ is the twist angle between the top- and bottom-layer spins of each CrSBr atomic unit cells (extracted from Supplementary Fig. 6 in SM), $N_{x,y}$ is the total number of CrSBr atomic unit cells within a moiré unit cell, and $\Delta E$ is the only fitting parameter with a fitted value of 12.5 meV for both 0.4°– and 1.1°–tBLs.

**References**
1. Mak, K. F. & Shan, J. Semiconductor moiré materials. *Nat. Nanotechnol.* **17**, 686–695 (2022).
2. Andrei, E. Y. *et al.* The marvels of moiré materials. *Nat Rev Mater* **6**, 201–206 (2021).
3. He, F. *et al.* Moiré Patterns in 2D Materials: A Review. *ACS Nano* **15**, 5944–5958 (2021).
4. Kennes, D. M. *et al.* Moiré heterostructures as a condensed-matter quantum simulator. *Nat. Phys.* **17**, 155–163 (2021).
5. Bernevig, B. A. *et al.* Fractional quantization in insulators from Hall to Chern. *Nat. Phys.* **21**, 1702–1713 (2025).
6. Seyler, K. L. *et al.* Signatures of moiré-trapped valley excitons in MoSe2/WSe2 heterobilayers. *Nature* **567**, 66–70 (2019).
7. Alexeev, E. M. *et al.* Resonantly hybridized excitons in moiré superlattices in van der Waals heterostructures. *Nature* **567**, 81–86 (2019).

8. Tran, K. *et al.* Evidence for moiré excitons in van der Waals heterostructures. *Nature* **567**, 71–75 (2019).
9. Jin, C. *et al.* Observation of moiré excitons in WSe2/WS2 heterostructure superlattices. *Nature* **567**, 76–80 (2019).
10. Zhang, L. *et al.* Van der Waals heterostructure polaritons with moiré-induced nonlinearity. *Nature* **591**, 61–65 (2021).
11. Yu, H., Liu, G.-B., Tang, J., Xu, X. & Yao, W. Moiré excitons: From programmable quantum emitter arrays to spin-orbit–coupled artificial lattices. *Sci. Adv.* **3**, e1701696 (2017).
12. Wang, X. *et al.* Light-induced ferromagnetism in moiré superlattices. *Nature* **604**, 468–473 (2022).
13. Xiong, R. *et al.* Correlated insulator of excitons in $WSe_2$/$WS_2$ moiré superlattices. *Science* **380**, 860–864 (2023).
14. Li, H. *et al.* Imaging moiré excited states with photocurrent tunnelling microscopy. *Nat. Mater.* **23**, 633–638 (2024).
15. Mellado, P. Magnetic Moiré Systems: a review. *J. Phys.: Condens. Matter* **37**, 323001 (2025).
16. Wu, M., Li, Z., Cao, T. & Louie, S. G. Physical origin of giant excitonic and magneto-optical responses in two-dimensional ferromagnetic insulators. *Nat Commun* **10**, 2371 (2019).
17. Jin, W. *et al.* Observation of the polaronic character of excitons in a two-dimensional semiconducting magnet CrI3. *Nat Commun* **11**, 4780 (2020).
18. Kang, S. *et al.* Coherent many-body exciton in van der Waals antiferromagnet NiPS3. *Nature* **583**, 785–789 (2020).
19. Hwangbo, K. *et al.* Highly anisotropic excitons and multiple phonon bound states in a van der Waals antiferromagnetic insulator. *Nat. Nanotechnol.* **16**, 655–660 (2021).
20. Ziebel, M. E. *et al.* CrSBr: An Air-Stable, Two-Dimensional Magnetic Semiconductor. *Nano Lett.* **24**, 4319–4329 (2024).
21. Dirnberger, F. *et al.* Magneto-optics in a van der Waals magnet tuned by self-hybridized polaritons. *Nature* **620**, 533–537 (2023).
22. Wang, T. *et al.* Magnetically-dressed CrSBr exciton-polaritons in ultrastrong coupling regime. *Nat Commun* **14**, 5966 (2023).
23. Li, Q. *et al.* Two-Dimensional Magnetic Exciton Polariton with Strongly Coupled Atomic and Photonic Anisotropies. *Phys. Rev. Lett.* **133**, 266901 (2024).
24. Shao, Y. *et al.* Magnetically confined surface and bulk excitons in a layered antiferromagnet. *Nat. Mater.* **24**, 391–398 (2025).
25. Smolenski, S. *et al.* Large exciton binding energy in a bulk van der Waals magnet from quasi-1D electronic localization. *Nat Commun* **16**, 1134 (2025).
26. Wilson, N. P. *et al.* Interlayer electronic coupling on demand in a 2D magnetic semiconductor. *Nat. Mater.* **20**, 1657–1662 (2021).
27. Bae, Y. J. *et al.* Exciton-coupled coherent magnons in a 2D semiconductor. *Nature* **609**, 282–286 (2022).
28. Diederich, G. M. *et al.* Tunable interaction between excitons and hybridized magnons in a layered semiconductor. *Nat. Nanotechnol.* **18**, 23–28 (2023).
29. Liu, J., Zhang, X. & Lu, G. Moiré magnetism and moiré excitons in twisted CrSBr bilayers. *Proc. Natl. Acad. Sci. U.S.A.* **122**, e2413326121 (2025).
30. Chen, Y. *et al.* Twist-assisted all-antiferromagnetic tunnel junction in the atomic limit. *Nature* **632**, 1045–1051 (2024).
31. Boix-Constant, C. *et al.* Multistep magnetization switching in orthogonally twisted ferromagnetic monolayers. *Nat. Mater.* **23**, 212–218 (2024).

32. Healey, A. J. *et al.* Imaging magnetic switching in orthogonally twisted stacks of a van der Waals antiferromagnet. arXiv.2410.19209 (2024).
33. Sun, Y. *et al.* Electron-magnon coupling at the interface of a 'twin-twisted' antiferromagnet. arXiv.2506.10080 (2025).
34. Boix-Constant, C. *et al.* Programmable Magnetic Hysteresis in Orthogonally-Twisted 2D CrSBr Magnets via Stacking Engineering. *Advanced Materials* **37**, 2415774 (2025).
35. Li, Q. *et al.* Twist Engineering of Anisotropic Excitonic and Optical Properties of a Two-Dimensional Magnetic Semiconductor. *Phys. Rev. Lett.* **135**, 156901 (2025).
36. Sung, S. H. *et al.* Torsional periodic lattice distortions and diffraction of twisted 2D materials. *Nat Commun* **13**, 7826 (2022).
37. Xie, H. *et al.* Evidence of non-collinear spin texture in magnetic moiré superlattices. *Nat. Phys.* **19**, 1150–1155 (2023).
38. Song, T. *et al.* Direct visualization of magnetic domains and moiré magnetism in twisted 2D magnets. *Science* **374**, 1140–1144 (2021).
39. Xu, Y. *et al.* Coexisting ferromagnetic–antiferromagnetic state in twisted bilayer CrI3. *Nat. Nanotechnol.* **17**, 143–147 (2022).
40. Xie, H. *et al.* Twist engineering of the two-dimensional magnetism in double bilayer chromium triiodide homostructures. *Nat. Phys.* **18**, 30–36 (2022).
41. Li, S. *et al.* Observation of stacking engineered magnetic phase transitions within moiré supercells of twisted van der Waals magnets. *Nat Commun* **15**, 5712 (2024).
42. Wang, C., Gao, Y., Lv, H., Xu, X. & Xiao, D. Stacking Domain Wall Magnons in Twisted van der Waals Magnets. *Phys. Rev. Lett.* **125**, 247201 (2020).
43. Mehio, O. *et al.* A Hubbard exciton fluid in a photo-doped antiferromagnetic Mott insulator. *Nat. Phys.* **19**, 1876–1882 (2023).
44. Liu, W. *et al.* A Three-Stage Magnetic Phase Transition Revealed in Ultrahigh-Quality van der Waals Bulk Magnet CrSBr. *ACS Nano* **16**, 15917–15926 (2022).
45. Carr, S. *et al.* Relaxation and domain formation in incommensurate two-dimensional heterostructures. *Phys. Rev. B* **98**, 224102 (2018).
46. Enkovaara, J. *et al.* Electronic structure calculations with GPAW: a real-space implementation of the projector augmented-wave method. *J. Phys.: Condens. Matter* **22**, 253202 (2010).
47. Mortensen, J. J. *et al.* GPAW: An open Python package for electronic structure calculations. *The Journal of Chemical Physics* **160**, 092503 (2024).
48. GiSser, O. & Paul, W. Magnetic properties of CrSBr. Magnetic properties of CrSBr. *Journal of Magnetism and Magnetic Materials* **92**, 129-136 (1990)
49. Larsen, A. H. *et al.* libvdwxc: a library for exchange–correlation functionals in the vdW-DF family. *Modelling Simul. Mater. Sci. Eng.* **25**, 065004 (2017).
50. Weston, A. *et al.* Atomic reconstruction in twisted bilayers of transition metal dichalcogenides. *Nat. Nanotechnol.* **15**, 592–597 (2020).
51. Enaldiev, V. V., Zólyomi, V., Yelgel, C., Magorrian, S. J. & Fal'ko, V. I. Stacking Domains and Dislocation Networks in Marginally Twisted Bilayers of Transition Metal Dichalcogenides. *Phys. Rev. Lett.* **124**, 206101 (2020).
52. Perdew, J. P., Burke, K. & Ernzerhof, M. Generalized Gradient Approximation Made Simple. *Phys. Rev. Lett.* **77**, 3865–3868 (1996).
53. Jochym, P. T. & Parlinski, K. Ab initio lattice dynamics and elastic constants of ZrC. *Eur. Phys. J. B* **15**, 265–268 (2000).
54. Beal, L. D. R., Hill, D. C., Martin, R. A. & Hedengren, J. D. GEKKO Optimization Suite. *Processes* **6**, 106 (2018).

55. Li, H. *et al.* Stacking effects on magnetic, vibrational, and optical properties of CrSBr bilayers. *Phys. Rev. B* **111**, 125411 (2025).
56. Giannozzi, P. *et al.* Advanced capabilities for materials modelling with Quantum ESPRESSO. *J. Phys.: Condens. Matter* **29**, 465901 (2017).
57. Liechtenstein, A. I., Anisimov, V. I. & Zaanen, J. Density-functional theory and strong interactions: Orbital ordering in Mott-Hubbard insulators. *Phys. Rev. B* **52**, R5467–R5470 (1995).
58. Magnes package. https://gitlab.com/alepoydes/magnes/-/tree/develop.
59. Shaban, P. S., Lobanov, I. S., Uzdin, V. M. & Iorsh, I. V. Skyrmion dynamics in moiré magnets. *Phys. Rev. B* **108**, 174440 (2023).
60. Abert, C. Micromagnetics and spintronics: models and numerical methods. *Eur. Phys. J. B* **92**, 120 (2019).
61. Yang, K., Wang, G., Liu, L., Lu, D. & Wu, H. Triaxial magnetic anisotropy in the two-dimensional ferromagnetic semiconductor CrSBr. *Phys. Rev. B* **104**, 144416 (2021).

**Acknowledgements**

Q.L. and H.D. acknowledge support by DARPA (HR0011-25-3-0317). H.D. acknowledge support by the Gordon and Betty Moore Foundation (GBMF10694), Office of Naval Research (N00014-21-1-2770) and Army Research Office (W911NF-25-1-0055). The work of A.S. and I.A.S. was supported by Priority 2030 Federal Academic Leadership Program. Computer resources and research IT are provided by UTS of the University of Iceland through the Icelandic research e-Infrastructure project (IREI), funded by the Icelandic Infrastructure Fund. R.H. acknowledges support from the U.S. Department of Energy (DOE), Office of Science, Basic Energy Science (BES), under award DE-SC0024147. The work conducted at University of Texas at Dallas was supported by US Air Force Office of Scientific Research Grant No. FA9550-19-1-0037, National Science Foundation- DMREF-2324033, and Office of Naval Research grant no. N00014-23-1-2020. L.Z. acknowledges the support by the U.S. Department of Energy (DOE), Office of Science, Basic Energy Science (BES), under award No. DE-SC0024145. C.R.D. acknowledges support from the U.S. Department of Energy (DOE), Office of Science, Basic Energy Sciences (BES), under award No. DE-SC0024870.

**Author contributions**

Q.L. and H.D. conceived the research. Q.L. performed the optical measurements. A.S., I.L., V.U., and I.A.S. performed the theoretical calculation and modelling. N.A. and R.H. performed the SAED and TEM measurements. A.A., Q.L., Y.Y., and W.A. fabricated CrSBr tBL samples. W.L., Z.Z., and B.L. provided the CrSBr single crystals. S.L., C.R.D., K.S., and L.Z. assisted with analysis. Q.L., A.S., and H.D. wrote the manuscript. All authors read and commented on the manuscript.

## Competing interests

The authors declare no competing interests.